\documentclass[nohyper,12pt,letterpaper]{JHEP}
\usepackage{epsfig}

\newcommand{\dd}{\mathrm{d}}
\newcommand{\dt}{\mathrm{dt}}
\newcommand{\dr}{\mathrm{dr}}
\newcommand{\ds}{\mathrm{ds}}

\newcommand\MM{\mathcal{M}}
\newcommand\Mp{\mathcal{M}_{+}}
\newcommand\Mm{\mathcal{M}_{-}}

\newcommand\si{\sigma}
\newcommand\Si{\Sigma}
\newcommand\ta{\tau}

\title{Entropy of the Stiffest Stars}

\author{T. Banks${}^{\dagger,\diamond}$, W. Fischler${}^{\star}$, A. Kashani-Poor${}^{\star}$, R. McNees${}^{\star}$, S. Paban${}^{\star}$ \\
  ${}^{\dagger}$ Department of Physics and Institute for Particle Physics\\
   University of California, Santa Cruz, CA 95064\\
\vspace{0.2cm}
 ${}^{\diamond}$ Department of Physics and Astronomy\\ 
  Rutgers University NHETC, Piscataway, NJ 08855 \\
E-mail: \email{banks@scipp.ucsc.edu} \\
\vspace{0.2cm}
   ${}^{\star}$ Department of Physics\\
   University of Texas, Austin, TX 78712\\
E-mail: \email{fischler, kashani, mcnees \\
\hspace{1.2cm}paban, @physics.utexas.edu}}

\abstract{We analyze the properties of stars whose interior
is described by the stiffest equation of state consistent with causality.
We note the remarkable fact that the entropy of such stars scales like 
the area.}


\received{???????? ?st, 2000} \accepted{???????? ?th, 1998}
\preprint{\hepth{0206096}\\RUNHETC-2002-20\\SCIPP-00/00\\UTTG-04-02}

\begin{document}

\section{\bf Introduction}

The paradigm of holography \cite{'tHooft:gx,Susskind:1994vu}  originally emerged from the study of
black  holes. Over  the  past  few years  string  theory has  produced
explicit  examples like the  AdS-CFT correspondence,  where holography
takes on a precise operational meaning.
The holographic principle is also realized in matrix theory, albeit in
a more  complicated fashion.  Examples  like these support  the belief
that holography might be a fundamental property of quantum gravity. If
this is  the case,  then it  is natural to  expect that  holography has
important consequences for cosmology. Recently Banks and Fischler developed a cosmology based on the holographic
principle \cite{Banks:2001yp}. Central to their discussion is a perfect fluid with equation of
state $p=\rho$. This equation of state represents the stiffest perfect fluid
consistent with causality. Using the holographic principle they argue for the
existence of a robust initial state for the universe which is dominated by such a
fluid. In \cite{Banks:2001px}, Banks and Fischler work out the consequences of such an initial
state for the universe, especially at late times.

The appearance of a $p=\rho$ fluid in the context of early cosmology is not completely unexpected when taking the connection of holography and causality into account. Since the $p=\rho$ equation of state describes the stiffest perfect fluid still consistent with causality, one might anticipate that it plays a similar demarcating role with regard to holography. In \cite{Fischler:1998st}, Fischler and Susskind demonstrate that this is indeed the case in an FRW universe: a $p=\rho$ fluid parametrically packs the most entropy within a given volume among all perfect fluids with linear equation of state. This property is reminiscent of black holes in static spacetimes.
In this paper, we are therefore lead to explore static solutions of Einstein's equations based on a $p=\rho$ fluid. Such solutions exist for any $p= \kappa \rho$ fluid. However, for all $\kappa$, the corresponding configurations extend out to infinity. Closing our eyes to this for an instant, we consider the entropy contained within a given radius and find indeed that the $p=\rho$ fluid parametrically packs more entropy than any other linear perfect fluid consistent with causality. What is more, beyond a certain minimal radius, the scale being set by the central pressure of the solution, the entropy within a radius for a $p=\rho$ solution scales like the surface area of a sphere of that radius.

The solutions described above are infinite in extent, with energy
density and pressure falling off asymptotically as $r^{-2}$. We have
two options if we wish to obtain finite size, star-like
configurations. The first, a rough analogy of a neutron star, is to
surround a $p=\rho$ core by an ideal fluid with a polytropic equation
of state. The pressure of the enveloping fluid drops to zero at a
finite radius, which we define as the surface of the object.
However, the mass and entropy of the polytrope invariably
dominate those of the core and obscure the scaling of the entropy
described above. The second option, which is more interesting for our
purposes, is to surround a $p=\rho$ core with a thin membrane.

This paper is organized as follows. In section \ref{properties}, we will review the properties of the $p=\rho$
perfect fluid. We then turn to the Tolman-Oppenheimer-Volkov (TOV) equation \cite{Tolman:1934,Oppenheimer:1939ne,Tolman:1939}
 for perfect fluid stars and discuss its properties in section \ref{TOV}.  We show that solutions of the TOV equation for a linear perfect fluid exhibit a
symmetry that determines how the total mass and entropy of the solution scale with
its size \cite{Bondi,Sorkin:1981}. We use this scaling behavior to determine the amount of entropy linear perfect fluids can carry within a ball of given radius.
In section \ref{membrane}, we discuss how to surround solutions of the TOV equation with a membrane. We show that the membrane can always be chosen such as not to impose any restrictions on the size of the object. We analyze the stability of these star-like objects in section \ref{stability}. We find a maximal radius of stability at given central pressure, and show that this radius is large enough to allow the onset of the area scaling behavior of the entropy of the $p=\rho$ solution. We end with some conclusions.

\section{\bf Properties of the  $p = \rho$ fluid}  \label{properties}

Holography and causality are closely intertwined. The covariant entropy bound due to Bousso (see \cite{Bousso:2002ju} and references therein) has causality built into its formulation, which is based on light cones. The first use of light cones in the context of holography by Fischler and Susskind was directly motivated by implementing causality in a generalized formulation of holography that would be valid in cosmology.
The expectation that a $p = \rho$ fluid should have interesting holographic properties is rooted in this observation, since, among all fluids with a linear equation of state, $p = \kappa \rho$, $\kappa=1$ has the stiffest equation of state
consistent with causality
\cite{Zeldovich:1961}. In this medium, the speed of sound is the speed of light in
vacuum. We thus may wonder whether this fluid has a similar demarcation role with regard to the holographic properties of perfect fluids.

The first indication that this might be the case was already found in \cite{Fischler:1998st} . The authors remarked there that in an FRW context with a homogeneous $p=\rho$ fluid, the entropy contained within a sphere of radius equal to the particle horizon size at a given time scales with the area of that sphere. For any other perfect fluid, the entropy scales with less than the area. In this paper, we consider the entropy contained in a sphere of $p=\rho$ material in a static, rather than an FRW spacetime, and find similar results. In the rest of this section, we will summarize further properties of the $p=\rho$ fluid that make it an interesting building block for cosmological models.

An indication that this fluid might be important at early times comes from its behavior in an FRW setting. The density of $p=\kappa \rho$ fluids scales as $\rho \sim a^{-3(1+\kappa)}$, $a(t)$ being the scale factor of the FRW universe. The $\kappa=1$ fluid thus has the fastest blueshift (redshift) of all fluids with a linear equation of state respecting causality, as we go
towards (away) from the big bang. If there is some matter with this equation of state, it therefore dominates at early times.

The statistical mechanics of this fluid is quite remarkable \cite{Banks:2001px,Banks:2001yp}.
Thermodynamics gives the temperature scaling of the entropy and the energy densities for a $p = \kappa\rho$ fluid as
\begin{eqnarray}
s &\sim& {T^{1\over\kappa}} \label{entdens} \\
\rho &\sim& T^{{\kappa + 1}\over\kappa} \label{rho} \,.
\end{eqnarray} 
These equations apply to any number of spatial dimensions and imply for 
the case $\kappa = 1$ that $\sigma\sim{T}$ and $\rho\sim{T^2}$. The thermodynamics of the $p=\rho$ fluid is thus
that of a
$1 + 1$ dimensional CFT. The constants of proportionality in the equations (\ref{entdens}) and (\ref{rho}) depend on the 
microscopics of this fluid. Banks and Fischler use this property of the $p=\rho$ fluid to model an initial state for the universe which incorporates a solution to the horizon problem.

Finally, the study of quantum fluctuations of the energy density in this background reveals that
they are scale invariant i.e. the amplitude of these fluctuations at 
horizon
crossing is independent of the mode number \cite{Banks:2001px}. This is another indication that a fluid with this equation of state might play a role in early cosmology. 

The scale invariance can be
illustrated by looking at the Schr\"odinger equation that determines 
the wave
function of a minimally coupled
  scalar, $\phi$, which mimics a graviton, in
an homogeneous $p = \rho$ background in D spacetime dimensions. Using 
conformal
time $\tau$ we get

\begin{eqnarray} \label{schrod}
i\dot\Psi(\{{\chi}_{\vec{k}}\},\tau) = \sum_{\vec{k}}
({{p_{\vec{k}}^2}\over2} + {{\vec{k}^2 \chi_{\vec{k}}^2}\over2}
+{{\chi_{\vec{k}}^2}\over 8\tau^2})\Psi(\{{\chi}_{\vec{k}}\},\tau) \,,
\end{eqnarray}
where $\chi_{\vec{k}} = {\tau}^{1\over2}\phi_{\vec{k}}$. Recall that by the definition of conformal time, the particle horizon at time $\tau$ is located at $r=\tau$. The Schr\"odinger equation (\ref{schrod}) separates into equations for the wave functions
$\Psi({\chi}_{\vec{k}},\tau)$ which only depend on a single momentum
mode
$\chi_{\vec{k}}$ , where $\prod_{\vec{k}}\Psi({\chi}_{\vec{k}},\tau) =
\Psi(\{{\chi}_{\vec{k}}\},\tau)$:

\begin{eqnarray} \label{schrod_k}
i\dot\Psi({\chi}_{\vec{k}},\tau) =
({{p_{\vec{k}}^2}\over2} + {{\vec{k}^2 \chi_{\vec{k}}^2}\over2}
+{{\chi_{\vec{k}}^2}\over 8\tau^2})\Psi({\chi}_{\vec{k}},\tau)\,.
\end{eqnarray}
This last equation is invariant under rescaling of
$\tau$ keeping
$k\tau$ and ${\chi_{\vec{k}}}\over{\tau}^{1\over2}$ fixed. This means 
that the
probability distribution $|\Psi({\phi}_{\vec{k}},t)|^2$, where $t\sim
{\tau}^{D-1\over{D-2}}$, is invariant when the wavenumber is the inverse 
horizon
size.

A similar argument for this scaling can be obtained from considering the
Green's function equation for a minimally coupled scalar field in the 
$p =
\rho$ background,

\begin{eqnarray} \label{scaleinv}
\ddot{\phi}_{\vec{k}}(t) + 
{1\over{t}}{\dot{\phi}_{\vec{k}}(t)} +
{{k^2}\over{t^{2\over{D-1}}}}{\phi}_{\vec{k}}(t) = {\delta(t)\over {t}} \,.
\end{eqnarray}

This equation exhibits again the invariance ``at horizon crossing", i.e. 
the
equation is invariant under rescaling of the wave number keeping
$k t^{D-2\over{D-1}}$ fixed.

\section{\bf The TOV equation for a $p = \rho$ star} \label{TOV}

In this section, we will demonstrate that the entropy of a $p=\rho$ configuration beyond a certain minimal size (the scale being set by the value of the central pressure) scales like the area of the sphere encompassing it. There are two ingredients to this discussion: the scaling properties of the TOV equation, and the existence of an analytic solution for the pressure \cite{Misner:1964} 
\begin{eqnarray} \label{analytic}
p=\frac{2}{4\pi(1 + 6\kappa +\kappa^2)}\frac{1}{r^2}  \,.
\end{eqnarray}
This solution is unphysical due to its singular behavior at the origin. The singularity is sufficiently mild however for the solution to exhibit finite mass and entropy within any given radius. What is more, the values for pressure, mass, and entropy of the solutions with finite central pressure asymptote to the respective values these quantities acquire for the singular solution.

We will start by reviewing the TOV equation. Given an equation of state, this equation yields the pressure profile of a static, spherically symmetric solution to Einstein's equation with matter content a perfect fluid. We will be interested in perfect fluids with linear equation of state, $p = \kappa\rho$, in $D$ spacetime dimensions. The TOV equation in this case proves to have very useful scaling properties.

We parametrize the metric as follows:
\begin{equation} \label{metric}
ds^2 = -A(r)dt^2 + B(r)dr^2 + r^2d\Omega^2  \,.
\end{equation}
Einstein's equations can then be written in terms of the metric
coefficients  $A$, $B$, the pressure $p$, and the energy density $\rho$ of 
the fluid,
\begin{eqnarray}
{(D-2)\over 2r}{\partial_{r}(AB)\over{AB^2}} &=& 8\pi (p + \rho) \nonumber \\
{(D-2)\over r^2}[ D -3 -\partial_{r}({r\over{B}}) +{4-D\over{B}}]& =& 16\pi\rho\nonumber \\
\partial_{r}p &=&-(p + \rho){\partial_{r}A\over{2A}} \label{einstein}\,.
\end{eqnarray}
These equations are invariant under a rescaling of $r$, $p$, and $\rho$,
\begin{eqnarray}
r&\rightarrow&\lambda r \,,\nonumber \\
p&\rightarrow&\lambda^{-2} p \,,\nonumber \\ 
\rho&\rightarrow&\lambda^{-2} \rho \label{trans} \,,
\end{eqnarray}
keeping $A$ and  $B$ 
fixed \cite{Bondi,Sorkin:1981}. Note that for the analytic solution, this rescaling leaves the boundary conditions on $p$ and $\rho$ at the origin fixed.

Once we specify an equation of state $p(\rho)$, we can obtain the density profile for a solution of equations (\ref{einstein}) by solving the TOV equation
\begin{eqnarray}
\frac{dp}{dr} &=& - (p + \rho) \frac{1}{2 r^{D-2}(1-\frac{c_D m}{r^{D-3}})}
\left(\frac{16\pi}{D-2}
r^{D-1} p + (D-3)c_D m\right)\,,
\end{eqnarray}
where $m(r) = \Omega_D \int^r_0 dr'
r'^{D-2} \rho(r') $, $c_D=\frac{16 \pi}{(D-2)\Omega_D}$, and $\Omega_D = \frac{2\pi^{\frac{D-1}{2}}}{\Gamma(\frac{D-1}{2})}$.

It is straightforward to find the scaling of the mass $m(r)$  
as a function of $r$,
\begin{eqnarray}
m(r \Lambda ) &=& \Lambda^{D-3} m(r)\,.
\end{eqnarray}

We now determine how the entropy scales under the transformations (\ref{trans}).
From thermodynamics, we obtain an equation relating the entropy density $s$ to the energy density
\begin{equation}
s \sim \rho^{1\over1+\kappa} \,. \label{ent}
\end{equation} 
The proportionality 
constant
depends on the microscopics of the fluid. It is the ignorance of this constant (which will be different for different $\kappa$ and possible different realizations of fluids of same $\kappa$) which forces the use of the word `parametrically' in all statements made below.
Using equation (\ref{ent}), the total entropy within a radius $R$ is given by
\begin{equation}
S(R) \sim\int^R_0 dr r^{D-2} B^{1\over2}(r)\rho(r)^{1\over1+\kappa} \,.
\end{equation}
By the scaling properties of 
the TOV
equation, we obtain
\begin{equation}
S(R) \sim R^{D-1-{2\over1+\kappa}} \,.
\end{equation}
Notice that for $\kappa = 1$, the entropy scales like 
the
area, $R^{D-2}\sim M^{D-2}$, in all dimensions. For all other cases, $\kappa<1$, the entropy
scales according to a power less than the area. 

While these scaling relations also hold for solutions with finite energy density at the origin, they here relate the entropy and radius of solutions with different central pressure. However, one can show numerically that the entropy of all solutions of the TOV equation, beyond a certain radius, is well approximated by the entropy of the analytical solution. Hence, beyond this radius, the scaling behavior also approximately holds for fixed central pressure in the non-singular case (see section \ref{stability} for further discussion).

As an aside, we can determine the temperature of the fluid at any point of the 
fluid by thermodynamics. We obtain
\begin{equation}
T(r) \sim r^{-\frac{2\kappa}{1 + \kappa}} \,.
\end{equation}
This result is
valid in all dimensions. Note that for $\kappa=1$ the temperature scales like the inverse radius. 
This is reminiscent of the Hawking temperature of a black hole, which scales with the inverse radius of the size of the black hole.

What we have shown in this section is that the $ p=\rho$ fluid parametrically packs the most entropy in a given volume among all perfect fluids with linear equation of state. In the next two sections, we study how this result carries over to the case of finite size configurations.

\section{The Membrane}  \label{membrane}

In the previous section we considered a spherically symmetric
configuration of  $p =\rho$ fluid extending out to infinity. We found
that the entropy contained within a sphere of a given radius scales
parametrically like the area of that sphere. In this section, we ask
whether we can cut off the configuration at a finite radius to produce
a compact object whose entropy scales with its surface area. In the
next section, we will consider the stability of such objects.

One means of constructing a finite $p=\rho$ object is by joining the
TOV solution, cut off at some fixed radius, onto the vacuum
Schwarzschild solution via a thin shell.  By letting the  thickness of
the shell go to zero we obtain a star-like object, surrounded by a
membrane, whose gross properties are dominated by the $p=\rho$
solution to the TOV equation.  The membrane adds a delta function
discontinuity to the stress-energy tensor
\begin{equation}
	T_{\mu\nu} = T_{\mu\nu}^{matter} + S_{\mu\nu} \, \delta(n) \,,
	\label{emt}
\end{equation}
where $n$ is a coordinate normal to the membrane that defines the unit
normal vector $n_\mu = \partial_\mu n$. The membrane stress-energy
tensor $S_{\mu\nu}$ is given by
\begin{equation}
	S_{\mu\nu} = (\sigma + \tau)\,U_{\mu}U_{\nu} +
	\tau\,h_{\mu\nu} \,,   \label{shell}
\end{equation}
with $\sigma$ the energy density of the membrane, $\tau$ the tension,
and $h_{\mu\nu}$ the induced metric on the shell. With these
conventions positive $\tau$ corresponds to gas-like behavior, with an
increase in the surface area of the membrane implying a decrease in
its internal energy, while negative $\tau$ indicates a true
membrane-like behavior where an increase in surface area gives an
increase in the internal energy of the membrane.

In the following, we will analyze Einstein's equations with a true membrane
(i.e. negative $\tau$) encompassing our TOV solution from the previous
section. Our main concern is twofold: whether a physically reasonable
choice of parameters $\sigma$ and $\tau$ places any restrictions on
the radius at which we can cut off our solution to the TOV equation,
and whether the shell contributes sufficiently to the entropy of the
solution to disrupt the area scaling behavior. To simplify the
analysis, we will assume that we are at large enough radius so that
the metric  is well approximated by the analytic solution, equation
(\ref{analytic}).

The membrane divides the spacetime $\MM$ into two regions, $\Mp$ and
$\Mm$. The TOV solution determines the metric inside the membrane, in
the $\Mm$ region, while the metric in the outer region, $\Mp$, is
simply the Schwarzschild metric. The parameters of these two metrics
are related to the properties of the membrane via the Israel junction
conditions \cite{Israel:rt}. These conditions are succinctly expressed
in terms of the membrane stress-energy tensor and the intrinsic
curvature $K_{\mu\nu}=-\frac{1}{2}g_{\mu\nu,n}$, $n$ being the normal
direction to the membrane, in the $\Mp$ and $\Mm$ regions:
\begin{equation} 
	8 \pi S_{\mu\nu} = h_{\mu\nu} \, [K] - [K_{\mu\nu}] \,.
\end{equation} 
Square brackets indicate the jump in a quantity across the membrane,
i.e. $[K] = K_+ - K_-$.

To evaluate the junction conditions we need to specify the metric on
$\Mp$ and $\Mm$. For the static, spherically symmetric case it is of
the form
\begin{equation}
    \ds_{\pm}^{2} = -A_{\pm}(r) \, \dt^{2} + B_{\pm}(r) \, \dr^{2} +
    r^{2} \, \dd \Omega^{2} \,. \label{metric2}
\end{equation}
The membrane is located at $r=R$. The functions $A_+$ and $B_+$ are
simply the vacuum Schwarzschild solution:
\begin{equation}
B_{+} = \left(1-\frac{2m_{+}}{r}\right)^{-1} \;\;,\;\;\; A_{+} =
1-\frac{2m_{+}}{r} \,. \label{b+}
\end{equation}
The mass parameter $m_+$ will be specified by the Israel junction
conditions.  We approximate the metric in the $\Mm$ region by the
analytic solution (\ref{analytic}),
\begin{equation} 
p = \rho = \frac{1}{16 \pi r^2} \,,
\end{equation}
with $0 \leq r < R$. The function $B_{-}$ takes the same form as
$B_{+}$ in (\ref{b+}), with $m_+$ replaced by $m_-$, where
\begin{equation}
	 m_{-} = \frac{1}{4} \; r  \,.
\end{equation}
The functions in the metric (\ref{metric2}) are then
\begin{equation}
B_{-} = 2 \;\;,\;\;\; A_{-} = A_0 \, r^2  \,.
\end{equation}
Continuity of the induced metric on $\Si$ implies $A_{+}(R) =
A_{-}(R)$, though the derivatives of $A_{\pm}$ are discontinuous at
$r=R$.

Given the metric (\ref{metric2}), the Israel junction conditions yield
two independent equations for the static configuration ($\dot{R} =
\ddot{R} = 0$):
\begin{eqnarray}
  4 \pi R \si & = & k_{-}-k_{+} \\ \label{ks} 8 \pi (\si + 2 \ta) & =
  & k_{+} \, \partial_{r} \log{A_{+}} - k_{-} \, \partial_{r}
  \log{A_{-}} \,,
\end{eqnarray}
where we have defined the variables $k_{\pm}$
\begin{equation}
	k_{\pm} = \frac{1}{\sqrt{B_{\pm}(R)}} =
	\sqrt{1-\frac{2m_{\pm}}{R}} \,.
\end{equation}
The first condition gives the Schwarzschild mass for the exterior
solution as a function of $R$ and $\si$:
\begin{equation}
    m_{+} = \frac{1}{4}\;R + 4\pi R^{\,2} \si \cdot
    \frac{\,1}{\sqrt{2}} - 8 \pi^{2} R^{\,3} \si^{\,2} \,.
    \label{outermass}
\end{equation}
For fixed $R$ this mass increases monotonically on the interval $\si
\in(0,\si_{max})$, where $\si_{max}$ is defined by
\begin{equation}
   \frac{\partial m_+}{\partial
   \si}\left|_{\si_{max}}\frac{}{}\right. = 0 \,.
\end{equation}
Solving this equation yields
\begin{equation}
	\si_{max} = \frac{1}{4\pi\sqrt{2}R} \,.
\end{equation}
As $\si \rightarrow \si_{max}$, the mass $m_+ \rightarrow R /2$, which
corresponds to the system forming a black hole and establishes an
upper bound on $\si$. Equation (\ref{outermass}) hence places no
restriction on our choice of $R$, provided that we choose $\si$
sufficiently small.

The upper bound on $\si$ allows us to establish the parametric
dominance of the entropy due to the $p=\rho$ fluid over the entropy of
the membrane. If we make the weak assumption that the entropy density
$s$ of the membrane is proportional to some positive power of the
energy density,
\begin{equation}
	s  \sim  \si^{\beta} \;\;\;\;\;\mbox{where}\;\;\;\;\; \beta  >  0\,,
\end{equation}
then the membrane contributes an entropy
\begin{equation} \label{membraneentropy}
	S_{membrane} \sim 4\pi R^2 \si^{\beta}
\end{equation}
to the total entropy of the system. Since $\si$ is bounded by
$\si_{max} \sim R^{-1}$, the entropy of a membrane always grows more
slowly than its area.

The second condition (\ref{ks}) fixes the ratio of $\ta$ and $\si$ to be
\begin{equation}
   \frac{\ta}{\si} = \frac{1}{4} \; \left( \frac{\sqrt{2}}{k_+} \; - 1
   - \frac{1}{1-\sqrt{2}\,k_+}\; \right)  \,.
\end{equation}
This ratio is significant, as its absolute value represents the square of the velocity
with which disturbances propagate on the membrane. It must therefore be
bounded by the speed of light.

\begin{figure}
\begin{center}
\epsfig{file=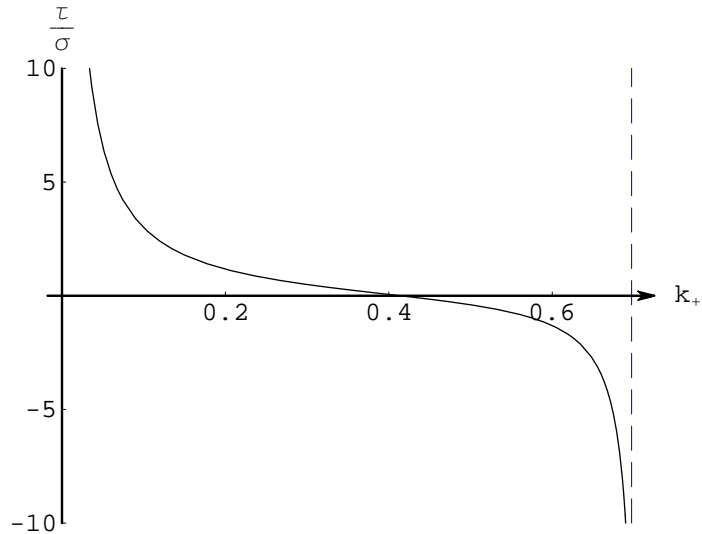}
\caption{\label{fig:toversvsk}The second Israel junction condition
gives the ratio $\ta_0 / \si_0$ as a function of the Schwarzschild
parameter in the metric on $\MM_{+}$.}
\end{center}
\end{figure}

By equation (\ref{ks}), $k_{+}$ is between $0$ and $k_{-} =
1/\sqrt{2}$. This ratio is shown in figure \ref{fig:toversvsk}. The
ratio $\ta / \si$ is zero at $k_+ = \sqrt{2}-1$. A membrane with $\si$
small but positive requires a negative tension. As we increase $\si$
the magnitude of this tension decreases, hitting $0$ at $\si \approx
.023/R$, then becoming positive and increasing monotonically until a
black hole is formed at $\si=\si_{max} \Rightarrow k_{+} = 0$.  From
figure \ref{fig:toversvsk}, we see that no restrictions arise on the
choice of the cutoff radius $R$, with the proviso that the radius is
large enough to justify the use of the analytic solution.

An alternative means of cutting off the TOV solution for a $p=\kappa
\rho$ fluid is by tacking on a polytrope at a chosen radius. For a
neutron star the transition from a linear equation of state (with
$\kappa=\frac{1}{3}$) to a polytrope occurs once the pressure has
fallen sufficiently under nuclear densities. A complication arises
when we try to mimic this transition for a $\kappa=1$ fluid. A polytrope
has equation of state $p \sim \rho^{\gamma}$, with $\gamma >1$. The
velocity of sound in this fluid is hence given by $v^2 =
\frac{p}{\rho}\gamma$. If we tack a polytrope on to a $p=\rho$ core, we
violate causality at the transition radius. This difficulty does not
arise in more realistic scenarios, where there is a smooth
interpolation between the linear and polytrope regimes. However, we
were not able to find a polytrope that cuts off our solution quickly
enough to not interfere with the area scaling of the entropy. Thus,
although we can construct stars with a $p=\rho$ core surrounded by
ordinary matter, they will not obey the area law. Furthermore, since
we find numerically that the polytrope radius must grow with the core
radius, such stars will have a maximum mass determined by properties
of ordinary matter. By contrast, the maximum mass of stiff stars
surrounded by a membrane is determined by microscopic properties of
the $p=\rho$ fluid, and by the central density. Nevertheless, with
remnants from early stages of the Banks-Fischler scenario in mind,
it would be interesting to study properties and possible observational
signatures of such polytropic $p=\rho$ stars.

\section{Stability}  \label{stability}

In this section, we perform a stability analysis of the TOV solution
encompassed by a membrane that we introduced above. As with ordinary
stars, we expect an instability to occur, at a given central pressure,
for radii exceeding a certain value. Our main concern is whether this
value is large enough to allow the onset of the area scaling behavior
for the entropy.

In the following, we follow closely the standard stability analysis
explained in \cite{Wheeler2} (see there for original references). The
novelty here is that due to the presence of the membrane, our energy
momentum tensor (see equation (\ref{emt})) contains delta function
contributions at the position of the membrane.

The starting point for the stability analysis is an equilibrium
solution to Einstein's equations, i.e. a solution to the TOV equation
with the equation of state $p=\rho$.   We consider a perturbation of
the equilibrium configuration $r \rightarrow r + \xi(r,t)$. After the
perturbation, the membrane is located at $R+\xi(R,t)$. Using the
equation of state and adiabaticity $\nabla \cdot (s u)=0$ with $s$ the
entropy density and $u$ the velocity 4-vector of the fluid, Einstein's
equations yield a differential equation for $\xi$ involving the
equilibrium configuration. Note that the only time dependence is
contained in the perturbation $\xi$, and that we can work in a gauge
in which the metric contains no mixed time and space
components. Hence, to first order in the perturbation and its
derivatives, the only time derivatives in the differential equation
for $\xi$ are of the form $\frac{\partial^2}{\partial t^2}\xi$ with a
time independent coefficient.  We can therefore make the separation
ansatz $\xi(r,t)=\xi_{\omega}(r) \exp (-i \omega t)$. $\omega^2$ is
necessarily real. For solutions $\xi_{\omega}$ with $\omega^2 > 0$, we
obtain two oscillatory modes, whereas for solutions with $\omega^2 <
0$, one mode dies off, the other increases exponentially, a sign of
instability.

The presence of the membrane comes in when determining the boundary
conditions that must be imposed on $\xi_{\omega}$. We bypass this
complication (to a certain extent, see end of section) by the
following reasoning, which allows us to focus attention on the bulk
\cite{Wheeler1}.  Solutions to the TOV equation can be parametrized by
the value of the energy density at the origin. As we change this value
smoothly, the values of $\omega^2$ as defined above vary smoothly as
well. Hence, a transition from oscillatory to unstable behavior must
occur via a zero mode $\omega^2 = 0$. A zero mode implies that two
static solutions to Einstein's equations are related to each other via
a small perturbation. We now introduce a quantity $X$ which among all
configurations with a given bulk entropy is maximized by those that
satisfy the TOV equation. Hence, two static solutions within a
perturbation of each other must both exhibit the same value of
$X$. This observation allows us, by finding the stationary points of
X, to determine at which central pressure the transition from a stable
to an unstable mode of configurations of a given fixed entropy occurs.

To introduce $X$, we follow \cite{Wheeler1} in parameterizing the mass
and radius of our configuration on a spatial hypersurface by an
entropy label $s$, i.e. $m(s)$ is the mass contained within the region
of radius $r(s)$ of total entropy $s$. An adiabatic perturbation of
the configuration now corresponds to perturbing $r(s)$, while keeping
the total entropy fixed. It is then not hard to show that the
resulting perturbation of the total mass is given by
\begin{eqnarray}
\delta M = - 4 \pi R^2 P \delta R + \mbox{term that vanishes when TOV
equation is satisfied.}
\end{eqnarray}

The quantity $X$ is just the total mass of the star as seen at
infinity when the pressure $P$ vanishes at the surface of the
star. For the case we are interested in, the pressure of the star is
finite at the outer bound of the TOV solution, and then is forced to
zero by the membrane. Hence, $X$ for us is determined by
\begin{eqnarray}
\delta X = \delta M + 4 \pi R^2 P \delta R \,. \label{deltaX}
\end{eqnarray}

Since we are interested in $X$ for solutions of TOV, we can rewrite
equation (\ref{deltaX}) by replacing $\delta$ by $\frac{d}{d \rho_c}$,
where $\rho_c$ denotes the density at the origin. We then obtain $X$,
up to a constant of integration, as
\begin{eqnarray}
X(\rho_c) = M(\rho_c) + 4\pi \int^{\rho_c} d\rho R(\rho)^2
P(\rho)\frac{d R}{d \rho} \,.  \label{X}
\end{eqnarray}

\begin{figure}
        \begin{center} \epsfig{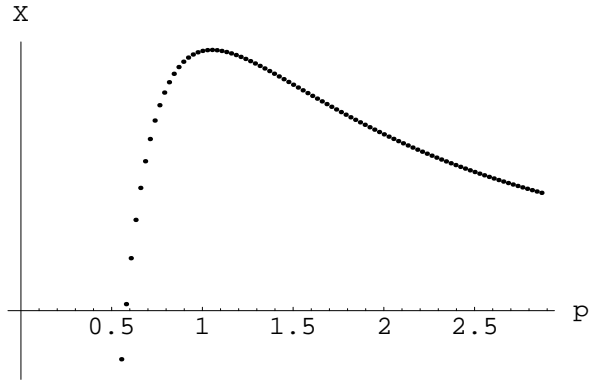}
         \caption{\label{fig:entfixed}X vs. central pressure
         (normalized such that the maximum occurs at $p=1$), keeping
         entropy fixed.}  \end{center} \end{figure}

We can now compute the integral in equation (\ref{X}) numerically and
plot $X$ vs. the central pressure at fixed entropy, see figure
\ref{fig:entfixed}.

\begin{figure}
        \begin{center} \epsfig{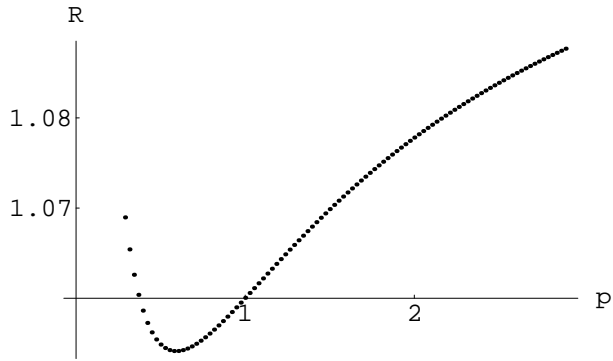}
         \caption{\label{fig:radvsden}Radius vs. central pressure
         (normalization from graph \ref{fig:entfixed}), keeping
         entropy fixed.}  \end{center} \end{figure}

Note that due to the scaling properties of the TOV equation, the first
extremal point of graph \ref{fig:entfixed} gives us the maximal radius
of stability for any central pressure we are interested in, given that
we establish the stability on the branch preceding this extremum. As
we can see from figure \ref{fig:radvsden}, the zero mode occurs for
$\rho =\frac{1}{\Lambda^2}$, and this corresponds to a configuration
with radius $R=1.06 \Lambda$, both in Planck units, where $\Lambda$ is
a scale factor at our disposal. Assume we are interested in a specific
value for the central pressure, and call this value $\rho_0$. Consider
a point $P=(\rho,R)$ on graph \ref{fig:radvsden} slightly preceding
the extremum. Choosing $\Lambda$ so that
$\rho_0=\frac{\rho}{\Lambda^2}$, we obtain the radius $R_0= \Lambda R$
for which we know that the configuration is stable. To see that this
is the maximal radius at which a configuration with central pressure
$\rho_0$ is stable, choose a slightly larger scale factor, s.t. the
density $\tilde{\rho}$ corresponding to a point
$\tilde{P}=(\tilde{\rho},\tilde{R})$ slightly past the extremum is
mapped to $\rho_0$. The radius $\tilde{R_0}=\tilde{\Lambda} \tilde{R}$
is slightly larger then $R_0$, and by choice of $\tilde{P}$
corresponds to an unstable configuration.

\begin{figure}
        \begin{center} \epsfig{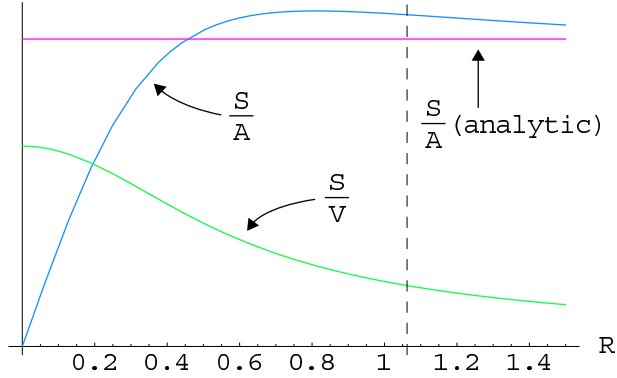}
         \caption{\label{fig:sovera}$\frac{S}{A}$ vs. radius and
         $\frac{S}{V}$ vs. radius, at fixed central pressure
         $p=1$. The transition from stability to instability in this
         plot occurs (with the choice of normalization of the axes) at
         the dashed line. $\frac{S}{A}$ for the analytic solution is given as
         reference.}  \end{center} \end{figure}

Finally, in figure \ref{fig:sovera}, we plot the entropy divided by
the area over the radius for central pressure $\rho
=\frac{1}{\Lambda^2}$. We can read off how well our solution is
approximated by the analytic solution (which exhibits exact area
scaling for the entropy) before we reach instability at $R=1.06
\Lambda$.

\begin{figure}
        \begin{center} \epsfig{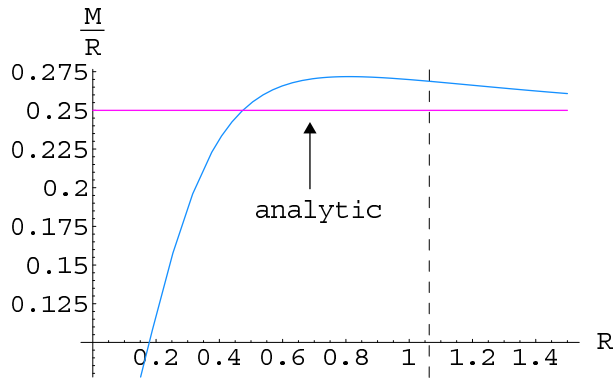}
         \caption{\label{fig:moverR} $\frac{m}{r}$ vs. radius at fixed
         central pressure.}  \end{center} \end{figure}

Note that for very weak gravity (measured by $\frac{m}{r} \ll 1$, see
figure (\ref{fig:moverR})), the entropy scales conventionally,
i.e. with the volume.

It remains to argue that the branch that precedes the first stationary
point depicted in figure \ref{fig:entfixed} corresponds to a region of
stability. Our argument on this account is merely suggestive: since
neutron stars have a $p=\kappa \rho$ core and are stable below a
central pressure of order $1/3$ of nuclear densities, such a stability
threshold should exist for our solution as well.  For certainty, one
can perform a full stability analysis as described in \cite{Wheeler2},
taking care however to choose test functions that are compatible with
the boundary conditions imposed by the presence of the membrane.

\section{Conclusion} 

$p=\rho$ is the stiffest equation of state compatible with
causality. In an FRW universe, a $p=\rho$ fluid is parametrically the
most entropic, given a volume, compatible with holography. We have
shown that this property prevails in static spacetimes.

We have considered three types of static, spherically symmetric
solutions based on this fluid. A solution with infinite pressure at
the origin and a naked curvature singularity exhibits exact area
scaling of its entropy. All other linear equations of state $p=\kappa
\rho$ consistent with causality have entropy scaling with less than
the area. Non-singular solutions extending out to infinity exist for
any finite central pressure. These asymptote to the singular solution,
and hence exhibit approximate area scaling of their entropy. Finally,
we have constructed finite-size solutions by cutting off the extended
solutions with a membrane. These non-singular star-like objects can
exhibit arbitrary radius at fixed central pressure. For large radius,
the entropy contribution of the membrane is negligible, such that
these objects exhibit approximate area scaling of their entropy. The
requirement of stability introduces a maximal radius at given central
pressure. The approximate area scaling law for the entropy sets in
before this bound is reached.

The message of this paper is twofold. First, we have witnessed a
compelling instance of the close connection between causality and
holography. It is intriguing to speculate whether these two basic
principles will ultimately prove to be two sides of the same
coin. Second, we have seen that a $p=\rho$ fluid not only can feature
reasonably as an initial state in cosmology but also leads to static
solutions entropically reminiscent of black holes. Perhaps this
equation of state catches, in some effective manner, a glimpse of what
may be the fundamental degrees of freedom of gravity.

\section{Acknowledgments}

We would like to thank Richard Matzner for a very useful
discussion.

The research of T.B was supported in part by DOE grant number
DE-FG03-92ER40689, the
research of W.F., A.K., R.M., and S.P. is based upon work supported by the National Science Foundation under Grant No. 0071512.


\newpage
\begin{thebibliography}{19}     

\bibitem{'tHooft:gx}
G.~'t Hooft,
``Dimensional Reduction In Quantum Gravity,''
arXiv:gr-qc/9310026.

\bibitem{Susskind:1994vu}
L.~Susskind,
``The World as a hologram,''
J.\ Math.\ Phys.\  {\bf 36}, 6377 (1995)
[arXiv:hep-th/9409089].


\bibitem{Banks:2001px}
T.~Banks and W.~Fischler,
``An holographic cosmology,''
arXiv:hep-th/0111142.

\bibitem{Banks:2001yp}
T.~Banks and W.~Fischler,
``M-theory observables for cosmological space-times,''
arXiv:hep-th/0102077.

\bibitem{Fischler:1998st}
W.~Fischler and L.~Susskind,
``Holography and cosmology,''
arXiv:hep-th/9806039.

\bibitem{Wheeler1}
B.~K.~Harrison, K.~S.~Thorne, M.~Wakano, J.~A.~Wheeler,
``Gravitation Theory and Gravitational Collapse,''
The University of Chicago Press, Chicago, 1965.

\bibitem{Bousso:2002ju}
R.~Bousso,
``The holographic principle,''
arXiv:hep-th/0203101.

\bibitem{Wheeler2}
C.~W.~Misner, K.~S.~Thorne, J.~A.~Wheeler,
``Gravitation,''
W.H. Freeman and Company, New York, 1970.

\bibitem{Oppenheimer:1939ne}
J.~R.~Oppenheimer and G.~M.~Volkoff,
``On Massive Neutron Cores,''
Phys.\ Rev.\  {\bf 55}, 374 (1939).

\bibitem{Tolman:1934}
R.~C.~Tolman,
``Relativity, Thermodynamics and Cosmology,''
Clarendon Press, Oxford, 1934.

\bibitem{Tolman:1939}
R.~C.~Tolman,
``Static Solutions to Einstein's Equations for Spheres of Fluid,''
Phys.\ Rev.\  {\bf 55}, 364 (1939).

\bibitem{Sorkin:1981}
R.~D.~Sorkin, R.~M.~Wald, Z.~Z.~Jiu,
``Entropy of Self-Gravitating Radiation,''
General Relativity and Gravitation, {\bf 13}, 1127 (1981).

\bibitem{Bondi}
H.~Bondi,
``Massive Spheres in General Relativity,''
Roy. Soc. London Proc. A, {\bf 282}, 303 (1964).


\bibitem{Misner:1964}
C.~W.~Misner and H.~S.~Zapolsky,
``High-Density Behavior and Dynamical Stability of Neutron Star Models,''
Phys.\ Rev.\ Lett.\ {\bf 12}, 635 (1964).


\bibitem{Zeldovich:1961}
Ya.~B.~Zel'dovich, Zh. Eksperim. i Teor. Fiz. {\bf 41}, 1609 (1961)
[translation: ``The Equation of State at Ultrahigh Densities and its Relativistic Limitations,'' Soviet Phys. -JETP {\bf 14}, 1143 (1962)].

\bibitem{Israel:rt}
W.~Israel,
``Singular Hypersurfaces And Thin Shells In General Relativity,''
Nuovo Cim.\ B {\bf 44S10}, 1 (1966)
[Erratum-ibid.\ B {\bf 48}, 463 (1967\ NUCIA,B44,1.1966)].

\end{thebibliography}
\end{document}